\documentclass[aps,prl,twocolumn,groupedaddress,superscriptaddress,reprint]{revtex4-1}

\usepackage{amsfonts} 
\usepackage{amssymb} 
\usepackage{graphicx}
\usepackage{amsmath}  
\usepackage{color}
\usepackage{ulem}
\newcommand{\pol}{\mathop{\mathrm{polylog}}}

\begin{document}

\title{Exact asymptotic behavior of magnetic stripe domain arrays}

\author{Tom H. Johansen}
\affiliation{Department of Physics, University of Oslo, 0316 Oslo, Norway}
\affiliation{Centre for Advanced Study at the Norwegian Academy of Science and Letters,
0271 Oslo, Norway}
\affiliation{Institute for Superconducting and Electronic Materials, 
University of Wollongong, Northfields Avenue, Wollongong, NSW 2522, 
Australia.}

\author{Alexey V. Pan}
\affiliation{Institute for Superconducting and Electronic Materials, 
University of Wollongong, Northfields Avenue, Wollongong, NSW 2522, 
Australia.}

\author{Yuri M. Galperin}
\affiliation{Department of Physics, University of Oslo, 0316 Oslo, Norway}
\affiliation{Centre for Advanced Study at the Norwegian Academy of Science and Letters,
0271 Oslo, Norway}
\affiliation{Physico-Technical Institute RAS, 194021 St. Petersburg, Russian Federation}
\affiliation{Argonne National Laboratory, 9700 S. Cass Ave., Lemont,  IL 60439, U. S. A.}

\date{\today}

\begin{abstract}
The classical problem of magnetic stripe domain behavior in films and plates with uniaxial 
magnetic anisotropy is addressed.  Exact analytical results are derived for the stripe 
domain widths as function 
of applied perpendicular field, $H$,  in the regime where the domain period becomes large. 
The stripe period diverges as $(H_c-H)^{-1/2}$, where $H_c$ is the critical (infinite period) field, 
an exact result  confirming a previous conjecture.
The magnetization approaches saturation as $(H_c-H)^{1/2}$, a behavior which 
compares excellently with experimental data obtained for a $4~\mu$m thick ferrite garnet film. 
The exact analytical solution provides a new basis for precise characterization of 
uniaxial magnetic films and plates, illustrated by a simple way to measure  the domain 
wall energy. The mathematical approach is applicable for similar analysis of a wide class of 
systems with competing interactions where a stripe domain phase is formed.

\end{abstract}

\maketitle

\vspace*{-5mm}

Systems with competing interactions, in particular those with short-range attractive and long-range 
repulsive interactions, commonly develop modulations in the 
order parameter and form domain structures often consisting of 
a stripe pattern~\cite{Seul, *Sagui}. 
Realizations are found in wide variety of systems, such as magnetic films and plates 
with uniaxial anisotropy~\cite{Kooy},
magnetic liquids~\cite{Flament, *Islam},  type-I superconductors in the intermediate 
state~\cite{Huebener, *Prozorov}, doped Mott
insulators~\cite{Tranquada95}, quantum Hall structures~\cite{Fogler96,*Koulakov96},
and monomolecular amphiphilic (``Langmuir")  films~\cite{deGennes82,*Gelbart96}. 

The uniaxial magnetic films, where ferrite garnets is a classical material studied extensively
 decades ago for use in bubble memory devices,~\cite{Bobeck book}
may be  regarded as a prototype system for stripe domain behavior.  Recently, 
the dynamical  behavior of the domains in thick garnet films showed a vast potential for 
manipulation of micron-sized superparamagnetic beads dispersed in a water layer
covering the  film. By applying magnetic fields with 
oscillating in- and out-of-plane components, new principles for micromachines like
colloidal ratchets, size separators, micro-tweezers and stirrers,  
etc. were demonstrated~\cite{Helseth0, *Helseth2, *TiernoPRL99, *TiernoPRL100}.
Moreover, it has been shown that the magnetic stripe domain structure, 
when placed adjacent to type-II superconductors, can strongly interact with the 
vortex matter, both in a manipulative way~\cite{Goa, *Helseth1, *Vestgaarden},  
and as a method to enhance flux pinning in the superconductor~\cite{Dao2011,*Iavarone2011,*Vlasko2012}.
Thus, one sees today considerable renewed interest in the collective behavior of 
magnetic stripe domains.

On  the theoretical side, the treatment of magnetic domains in plates with perpendicular easy-axis 
anisotropy placed in an external magnetic field is challenging.
Even solving the magnetostatic problem of one isolated linear stripe  surrounded by
 reverse magnetization
turned out rather complicated analytically, and for a regular array of alternating stripes
results were so far obtained only by numerical calculations~\cite{Cape, *Babcock}. 
In this work, based on the 
wall-energy model~\cite{Kooy},  i.e., assuming 
domains separated by infinitely thin walls oriented normal to the plate, we derive an exact analytical solution
for the behavior of a periodic array of interacting stripe domains  
in  increasing applied field. 
\begin{figure}[b!!]
\vspace*{-3mm}
\begin{center}
  \includegraphics[width=7.3cm]{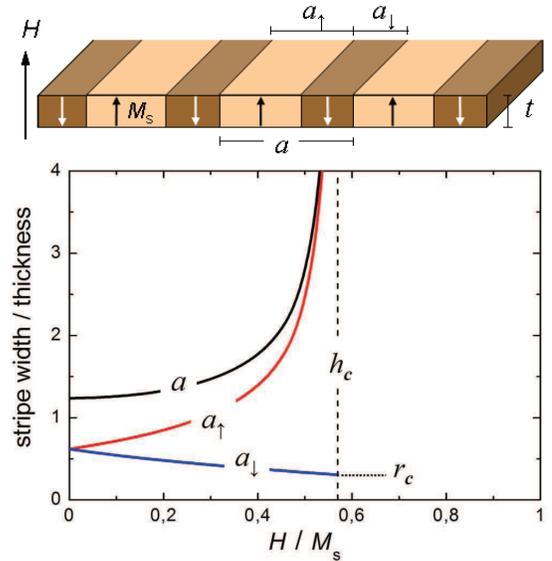}
\vspace*{-6mm}
\end{center}
\caption{Cross-section of a plate with magnetic stripe domain pattern (top),
  and numerical results for the domain widths as function
  of applied field for the case $\Lambda = 0.05$\label{Fig-1} (bottom).}
\end{figure}

Consider a uniaxial  plate of arbitrary thickness, $t$, where magnetic domains form a 
periodic lattice of parallel stripes
  with alternating magnetization $\pm M_s$, see Fig.~\ref{Fig-1}.  In an applied
perpendicular field, $H$, the domains magnetized parallel and
antiparallel to the field are characterized by their respective widths
$a_{\uparrow}$ and $a_{\downarrow}$, and 
the magnetization of the plate is 
$ M = M_s (a_{\uparrow} -a_{\downarrow})/a$, where $a=
a_{\uparrow} + a_{\downarrow}$ is the period of the stripe lattice.

Following the analysis of Kooy and Enz~\cite{Kooy}, the
energy density has three contributions; (\textit{i}) the cost of forming
 domain walls, characterized by the energy $\sigma_w$ per unit
wall area,  (\textit{ii})  the energy gain of aligning the magnetization
with the applied field, $-\mu_0 HM$, and  (\textit{iii}) the self-energy
of the domain structure (demagnetization energy). The total energy,
$U$, per unit volume of the plate can then be written
as~\cite{Bobeck book}
\begin{eqnarray}
\frac{U(m,a)}{\mu_0 M_s^2} &= & 2\Lambda \frac{t}{a} -mh +
\frac{1}{2}m^2 \nonumber \\
&& \hspace*{-13mm} + \, 2\frac{a}{t} \, \sum_{n=1}^\infty
\frac{\sin^2 [n \pi (1+m)/2]}{(n\pi)^3} \ (1-{\rm e}^{-n 2\pi t/a})
\, .
\end{eqnarray}
Here $m = M/M_s$, $h = H/M_s$, and $\Lambda = \lambda/t$ where
$\lambda = \sigma_w/\mu_0 M_s^2$ is a characteristic length.
The equilibrium magnetization and stripe period at a given applied field
is given by, $\partial U/\partial m = \partial U/\partial a = 0$, 
and expressed by the two equations,
\begin{eqnarray}
1-2x -\frac{2}{\pi} \mathcal{F}(2 \pi x, 2 \pi y) &=& h \, , \label{eq2} \\
  \mathcal{G}(2 \pi x, 2 \pi y) &=& 2 \pi  \Lambda \, . \label{eq3}
\end{eqnarray}
Here $x \equiv a_{\downarrow}/a = (1-m)/2 \, $   and $y \equiv
t/a \,$,  and 
\begin{eqnarray}
\mathcal{F}(x,y) &\equiv& \sum_{n=1}^\infty \frac{\sin nx}{n^2} \,
\frac{1- e^{-ny}}{y} \, ,  \label{d:005} \\
\mathcal{G}(x,y) &\equiv & 8\, \sum_{n=1}^\infty \frac{\sin^2 (n x/2)}{n^3} \,
\frac{1-(1+ny)e^{-ny}}{y^2}\, . \label{d:007}
\end{eqnarray}
The particular choice of new variables is motivated by the
numerical solution of the problem 
shown graphically in the lower panel of Fig.~\ref{Fig-1}
only for reference. The striking feature is that when the
applied field approaches a critical value, $H_c = h_c M_s$ where
$h_c <1$, both the period, $a$, of the domain lattice, and the
width, $a_{\uparrow}$, of the domains magnetized parallel to the
field will diverge, whereas the reverse domains contract only
moderately and terminate at a finite width $a_{\downarrow c}$. At
fields above $ H_c$, the material remains single-domain. In the new
variables the approach towards the critical values corresponds to
both $x, y \rightarrow 0$, while the ratio
$
r \equiv x/y = a_{\downarrow}/t
$
remains finite~\cite{comment}.

Focus of the present analysis is to determine analytically the
exact behavior as the field approaches the  critical value. 
We first derive the
relation between $h_c$ and $a_{\downarrow c}$.  For this, we
introduce the auxiliary function 
$p_k(z)  \equiv \sum_{n=1}^{\infty} n^{-(k+1)} \, {\rm e}^{-nz} =\pol (k,-z) \, ,$
and write
\begin{equation}
\mathcal{F}(x,y) = {\rm Im} [p_1(-i x)-p_1( y- ix) ]/ y \, .
\end{equation}
For small $|z|$  one has
\begin{equation}
p_1(z) = 
 (\pi^2/6) + z(\ln z -1) - (z^2/4) + (z^3/72) +\ldots ,
\end{equation}
which results in the following series expansion,
\begin{eqnarray}\label{d:006}
\mathcal{F}(x,y) &=&  \arctan (\frac {x}{y}) +
\frac{x}{y}\, \ln \sqrt{1+\frac{y^2}{x^2}}\, -\frac{x}{2}  \nonumber \\
  && + \frac{xy}{24} + \frac{2yx^3-xy^3}{2880} +\ldots .
\end{eqnarray}
Inserted in Eq.~(\ref{eq2})  it takes the form
\begin{eqnarray}
h & = & 1 - \frac{2}{\pi} \left[ \arctan(r) +
r \ln \sqrt{1+ r^{-2}} \right]   \nonumber \\
& & - \frac{\pi}{3r}x^2 + \frac{\pi^3}{90r}(2 - r^{-2})x^4 + \ldots
\, . \label{d006a}
\end{eqnarray}
The critical field is therefore given by
\begin{equation} \label{eq10}
h_c = \frac{2}{\pi} \arctan(r_c^{-1}) -  \frac{r_c}{\pi} \ln (1+
r_c^{-2}) \, ,
\end{equation}
where $r_c =a_{\downarrow c}/t$ is the critical, i.~e., the  terminal width of the
minority (anti-parallel to $\mathbf{H}$)  domains.
\begin{figure}[b!!]
\begin{center}
  \includegraphics[width=7.5cm]{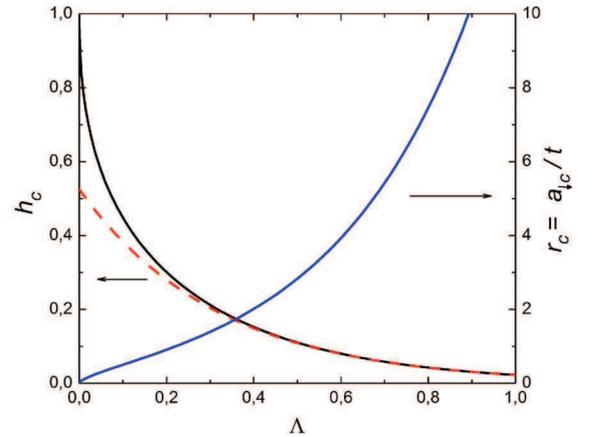}
\end{center}
\vspace*{-3mm}
  \caption{The critical field, $ h_c = H_c/M_s$, 
and terminal minority domain width, $r_c$,  
as functions of  $\Lambda = \sigma_w/(\mu_0 M_s^2t)$. 
The dashed line represents 
 $h_c(\Lambda) = (\sqrt{\rm e}/\pi) \, {\rm exp}(-\pi \Lambda)$ \label{fig2}}
\end{figure}

To find a relation between $r_c$ and the material parameter $\Lambda$,
a similar treatment is given to Eq.~(\ref{eq3}), using that
$\mathcal{G}(x,y)$ can be expressed as a combination of the real
parts of both $p_1(z)$ and $p_2(z)$ with complex arguments like
those in Eq.~(6). For small $|z|$, one has 
\begin{equation}
p_2(z)
=  \zeta(3) - \frac{\pi^2z}{6} + \frac{(3-2\ln z)z^2}{4} +
\frac{z^3}{12} -\frac{z^4}{288} + \ldots
\end{equation}
where $\zeta(n)$ is the Riemann zeta-function, and Eq.~(3) becomes
\begin{eqnarray}
 \ln(1+r^2) + r^2 \ln(1+r^{-2})
-\frac{\pi^2}{3}x^2 & & \nonumber \\
- \frac{\pi^4}{270}(3 -9r^{-2}+ r^{-4})x^4+ \ldots & = & 2 \pi
\Lambda \ . \label{d0011a}
\end{eqnarray}
The terminal width of the minority domains is therefore given by
\begin{equation} \label{eq13}
 \ln ( 1+ r_c^2 ) +
r_c^2  \ln (1+ r_c^{-2} )= 2 \pi \Lambda \, ,
\end{equation}
and is shown graphically in Fig.~\ref{fig2}. The figure also shows the dependence 
$h_c (\Lambda)$, which follows from  Eqs.~(10) and (13).
Both these curves, if replotted as functions of $\Lambda^{-1}$,  
agree excellently with the numerical solutions presented in Fig.~7 of the Ref.~\onlinecite{Cape}.
Note that for any material, i.~e., given $\sigma_w$ and $M_s$,
 the critical field decreases with $\Lambda$, and 
for $\Lambda > 0.2$ this dependence rapidly
approaches $h_c = (\sqrt{\rm e}/\pi) \,
{\rm exp}(-\pi \Lambda)$, shown as a dashed line in Fig.~2.

Consider next the behavior in the vicinity of $h=h_c$.
Expanding the Eqs.~(9) and (12) in Taylor series around the critical point
one finds to the lowest order
that $ \pi x^2 = 2r_c (h_c - h) $. It then follows that the stripe
pattern period, $a/t = r/x$, diverges according to
\begin{equation}
\frac{a}{t} =  \sqrt{\frac{\pi r_c}{2}} \ (h_c - h)^{-1/2} \, .
\end{equation}
At the same time, the reverse domain approaches its terminal width
as
\begin{equation}
\frac{a_{\downarrow}}{t} =  r_c + \frac{\pi(h_c -h)}{3 \,
\ln(1+r_c^{-2})}
 \, ,
\end{equation}
and the magnetization, $m = 1 - 2rt/a$, approaches saturation 
according to
\begin{equation}
m = 1 - \sqrt{\frac{8 r_c}{\pi}} \ \left( h_c - h \right)^{1/2} \,
\end{equation}
\begin{figure}[t!!]
\begin{center}
  \includegraphics[width=8.0cm]{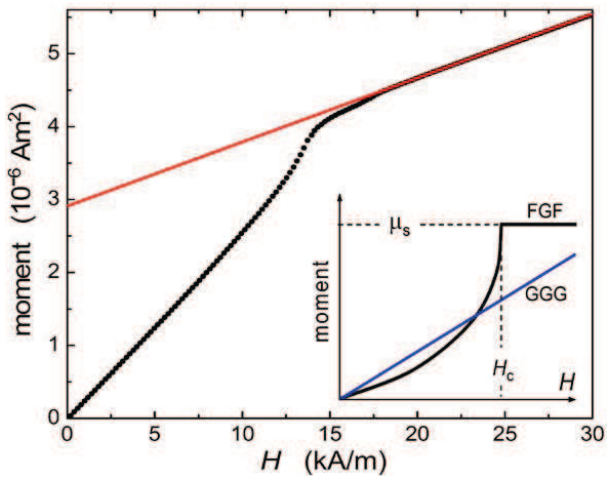}
\end{center}
  \caption{Magnetic moment of a ferrite garnet film versus applied perpendicular field. 
The straight line represents the contribution from the paramagnetic GGG substrate.
The inset shows a schematic of the two contributions to the moment.\label{Fig-3} }
\end{figure}
\begin{figure}[tt!!]
\begin{center}
  \includegraphics[width=8.4cm]{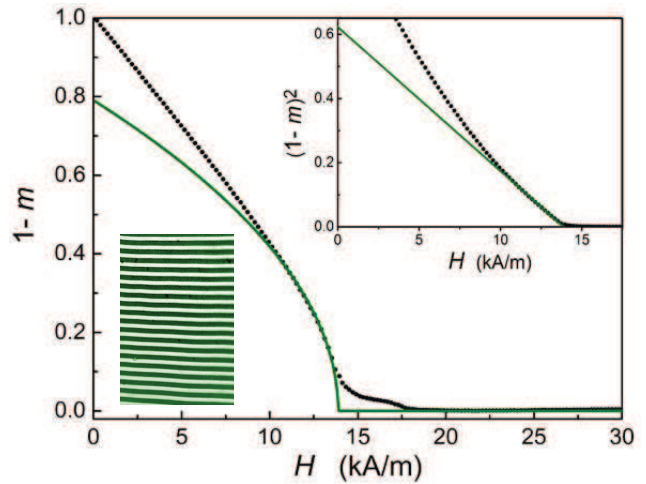}
\end{center}
\vspace*{-2mm}
  \caption{Reduced magnetization $(M_s -M)/M_s$, experimental data
  and fitted model behavior, Eq.(13),
   versus applied magnetic field.
 Upper  inset: The linear fit to the reduced magnetization squared. Lower inset: 
Magneto-optical image showing  that the FGF sample displays a parallel stripe domain 
pattern.}
\end{figure}

To compare the analytical results with the quantitative behavior of a typical
sample with magnetic stripe domains, we prepared a film of
bismuth-substituted ferrite garnet, (Y,Lu,Bi)$_3$(FeGa)$_5$O$_{12}$,
by liquid phase epitaxial growth on a  (111) oriented gadolinium
gallium garnet (GGG) substrate. Oxide powders of the constituent
rare earths, bismuth, iron and gallium, as well as PbO and
B$_2$O$_3$, were initially melted in a thick-walled platinum
crucible. To ensure homogeneity of the solution a stirrer mixed the
melt while being kept in the 3-zone resistive furnace at
1050~$^{\circ}$C for 30 minutes. Prior to the film growth the melt
temperature was reduced to 700~$^{\circ}$C.
The GGG wafer was mounted horizontally in a 3-finger
platinum holder attached to a shaft rotating by 60 rpm, and brought
slowly down towards the melt. Finally, the substrate was dipped into
the  melted for 8 minutes resulting in a macroscopically uniform ferrite garnet film
(FGF) grown on one side of the substrate. 
A nearly square plate of area $A= 21$~mm$^2$ 
was selected for measurements. The thickness of the FGF 
was determined by viewing the sample edge-on in a scanning electron 
microscope,  where a sharp contrast between the film and the substrate 
becomes visible. The ferrite garnet  thickness was $t = (4.0 \pm 0.2)~\mu$m.

Shown in Fig.~3 is the result of dc-magnetization measurements performed 
using a Quantum Design Magnetic Property Measurement System (SQUID magnetometer)  
with the FGF mounted perpendicular to the applied field. 
Above the field of $17$~kA/m, the data show linear increase, which is due to the 
para\-magnetic substrate, in combination with the FGF being single 
domain having a constant moment.
The fitted straight line intersects the vertical axis at a point which
determines the saturation  moment of the FGF sample,  
$\mu_s = 2.91~ 10^{-6}$~Am$^2$, which corresponds to
 $M_s = 34.6 $~kA/m. 

Figure~4 shows the reduced magnetic moment of the FGF obtained 
by subtracting  the paramagnetic background from the raw data.  
Based on the model result, Eq.~(16),  the reduced magnetization
was fitted to the predicted asymptotic form $(H_c - H)^{1/2}$ using data over a 
field range below the point where the moment saturates.
The best linear fit of the reduced moment squared is seen in the upper inset,
from which we find a critical field of $H_c = 13.9$~kA/m, and thus $h_c = 0.40$.
It  follows then  from Eq.~(\ref{eq10}) that  $r_c = 0.605$, 
and from Eq.~(\ref{eq13}) one finds $\Lambda = 0.126 $, and $\lambda = 0.126t=0.504$ microns. 
The specific wall energy has therefore the value $\sigma_w = 7.58\cdot 10^{-4}$ J/m$^2$.

In  previous analyses of the stripe domain problem, 
see e.g., Ref.~\onlinecite{Babcock}, it was suggested 
that as the applied field approaches  $h_c$, 
the stripe period diverges with a power $\beta \approx 0.5$.
In this work it has been shown that  $\beta = 1/2$ is an exact result.
Consider next what is the field range over which the asymptotic 
behavior is expected to be observed.
There are previous works~\cite{Cape} where  experimental data were 
fitted by numerical  $M-H$ curves
approaching  saturation seemingly with a  finite slope. 
Furthermore, in the classical book Ref.~\onlinecite{Bobeck book},  
the Fig.~2.3  shows  $M-H$ curves approaching saturation with a finite slope 
strongly depending on the sample thickness.
To resolve this apparent inconsistency, 
we analyzed the Eqs.~(9) and (12) up to the next order, i.~e.,  expanding 
them to $(r-r_c)^2$ and keeping the terms $\propto x^4$. 
The analysis shows that Eqs.~(14)--(16) provide a good description as long 
as 
\begin{equation}
 (h_c - h) \lesssim \min\{r_c,1\} \ .
\end{equation}
Thus, for thick plates, $t \gg \lambda$, one has a very 
small $2\pi \Lambda $ and  $r_c \approx [2\pi \Lambda/\ln(1/2\pi \Lambda)]^{1/2} $, 
which is much less than unity. Thus, it follows that 
the asymptotic  behavior (16) will be observed only very close to $h_c$. In practice, 
the field interval may be beyond experimental resolution, and the slope 
of the magnetization curve near $h = h_c$ appears finite. 
For thinner plates and films, $r_c$ rapidly increases, see Fig.~2, and the 
inequality (17)  becomes much weaker, and the range where one should observe 
the critical behavior  Eq.~(16)  will be sizable, as demonstrated in the 
present experiments. 

Finally, note the presence of a small shoulder in the reduced magnetization data in Fig.~4
seen just above $H_c $. 
Here the behavior deviates significantly from the model prediction, and is 
caused by topological fluctuations in the domain pattern. The role of
material  defects causing wall pinning becomes significant, and the stripes lose
 their alignment, a scenario which is readily seen visually by following the behavior 
using magneto-optical imaging~\cite{MOI}.

In summary, we have presented an analytical asymptotic solution to the 
problem of modeling the behavior of an infinite array of parallel alternating 
magnetic stripe domains subjected to a transverse field.
The mathematical approach used in this work can be applied to 
derive exact results also for other systems~\cite{Flament, GoldsteinPRL,*GoldsteinPRB}
 where stripe domain phases are 
formed and described by a configurational energy term similar to the one 
treated in the present case. 

The work was financially supported by the Australian Research Council 
International Linkage Project LX0990073 and Discovery Project DP0879933, 
and the Norwegian Research Council. 
YG is grateful to I.~Lukyanchuk and A.~Mel'nikov for discussions.


\begin{thebibliography}{99}
\bibitem{comment}
From the Eqs.~(2) and (3) it is not evident that such a critical
point should exist. Our choice of variables was made by postulating
an $h_c$ where $d$ diverges and $a_\downarrow$ remains finite (as
suggested by numerics), and realizing that the leading terms in the
analytical functions representing the Fourier sums depend only on
$x/y$ as $x, y \rightarrow 0$.
\end{thebibliography}
\end{document}